\documentstyle[pra,aps]{revtex}
\begin{document}
\twocolumn[\hsize\textwidth\columnwidth\hsize\csname@twocolumnfalse\endcsname

%\draft
\title{Small-amplitude normal modes of a
 vortex in a trapped Bose-Einstein condensate}
\author{Marion Linn$^{1,2}$ and Alexander L.~Fetter$^1$}
\address{$^1$Department of Physics, Stanford
University, Stanford, CA 94305-4060 \\ $^2$Physikalisches
Institut, Universit\"at Bonn, Nu\ss allee 12, D-53115 Bonn,
Germany}

\date{\today}

\maketitle

\begin{abstract}

We consider a cylindrically symmetric trap containing a small
Bose-Einstein condensate with a singly quantized vortex on the
axis of symmetry. A time-dependent variational Lagrangian analysis
yields the small-amplitude dynamics of the vortex and the
condensate, directly determining the equations of motion of the
coupled normal modes. As found previously from the Bogoliubov
equations, there are two rigid dipole modes and one anomalous mode
with a negative frequency when seen in the laboratory frame. \\

PACS number(s): 03.75.F, 05.30.Jp, 32.80.Pj, 67.40.Db\\

\end{abstract}
]

\tighten

\section{INTRODUCTION}

After the achievement of Bose-Einstein condensates in trapped
atomic gases ~\cite{And,Dav,Brad}, interest in these systems has
increased dramatically. An especially intriguing question concerns
condensates containing a vortex. The low-energy normal modes are
important in characterizing the behavior of such a system. They
have been studied extensively
~\cite{Dodd,Fet98,SF98,Svid98,Zam98,IsoJ99,Ga99}, mostly with the
Bogoliubov equations~\cite{Bog47} that involve quantum-mechanical
normal-mode amplitudes. Consequently, the associated dynamical
motion of the vortex core and the condensate center of mass can
only be obtained indirectly. Here, we use a variational Lagrangian
procedure to provide a more direct and intuitive treatment of the
small-amplitude normal modes of a small (weakly interacting)
Bose-Einstein condensate containing a singly quantized vortex. As
in previous applications~\cite{Svid98,Pe97}, this approach yields
a clear physical picture of the dynamics. We show that it
reproduces the three lowest modes as found from the more intricate
Bogoliubov analysis, thereby clarifying the physical
interpretation. This description is particularly interesting for
the anomalous mode that causes the local instability of the vortex
in a trap at rest~\cite{Dodd,Fet98,IsoJ99}.

The article is organized as follows: in Sec. II we derive the
effective Lagrangian for a slightly off-center vortex in a small
condensate. In the next section we solve the corresponding
equations of motion for the vortex and the condensate and discuss
the physical interpretation. Finally we summarize our results in
the conclusion.

\section{Effective Lagrangian}

We consider a small Bose-Einstein condensate with $N$ particles
containing a singly quantized vortex on the axis of symmetry, and
observed in a frame rotating with an angular velocity
$\Omega$.
The harmonic trap is axisymmetric, with radial and axial
frequencies $\omega_\perp$ and $\omega_z$ (the ratio
$\lambda=\omega_z/\omega_\perp$ characterizes the axial
asymmetry). The $s$-wave scattering length $a$ (assumed positive)
characterizes the interparticle interactions. For a general
condensate wave function $\Psi$, the Lagrangian for such a system
is given by
\begin{eqnarray}\label{leff}
  L(\Psi) = \int dV
  &\Big[& \frac{i}{2}\left( \Psi^* \partial_t \Psi
  -\Psi \partial_t \Psi^*\right) \nonumber \\
  & -& \Psi^* \left(H_0 -\Omega L_z\right)\Psi
  - 2 \pi \gamma |\Psi|^4 \Big] \, ,
\end{eqnarray}
where $H_0$ is the Hamiltonian for the noninteracting condensate
\begin{equation}\label{H0}
  H_0 = \frac{1}{2} \Bigg[ - \partial^2_x- \partial^2_y
+(x^2 +y^2) + \lambda \left( -\partial^2_z+z^2\right)\Bigg] \, ,
\end{equation}
and $L_z = -i(x \partial_y - y \partial_x)$ is the $z$-component
of the angular-momentum operator. All quantities are expressed in
dimensionless units (the radial and axial coordinates are scaled
with the radial and axial  oscillator lengths $d_{\perp} =
\sqrt{\hbar/M \omega_{\perp}}$ and $d_z = \sqrt{\hbar/M
\omega_z}$, and frequencies are scaled with the radial trap
frequency $\omega_\perp$). The condensate wave function $\Psi$ is
normalized to unity and $\gamma \equiv  Na/d_z$ is the small
interaction parameter.

Next we construct a trial condensate wave function. The condensate
wave function is assumed to be unchanged with respect to the
noninteracting case along the axis of symmetry, which allows us to
use the ground-state gaussian $\varphi_0(z)
=\pi^{-1/4}\,\exp{(-z^2/2)}$. The radial wave function is based on
the vortex ground state, namely a two-dimensional gaussian times
the vortex factor $x + i y = r e^{i \varphi}$. We are interested
in the relative motion of the vortex which requires introducing
time-dependent parameters for the vortex position. In the
weak-coupling limit, the radius of the vortex core is comparable
to the radius of the condensate, so that the displacement ${\bf
r}_0 (t)=[x_0 (t),y_0(t)]$ of the vortex, the displacement ${\bf
r}_1 (t)=[x_1 (t),y_1 (t)]$ of the condensate, and the induced
velocity of the condensate $\bbox{\alpha} (t) =[\alpha_x
(t),\alpha_y (t)]$ must all be included. Hence we use the
following trial function
\begin{eqnarray} \label{trial}
  \Psi_v(x,y,z,t)&=& \frac{C}{{\pi}^{3/4}}
  [(x-x_0)+i(y-y_0)]\nonumber \\
  &\times& e^{-{1 \over 2} \left[ (x-x_1)^2
  +(y-y_1)^2+z^2\right]}e^{i(\alpha_x x+\alpha_y y)}
  \, ,
\end{eqnarray}
where $C^{-2} = 1+|{\bf r}_1-{\bf r}_0|^2$. Since we focus on the
dynamical motion of the vortex and the condensate, we ignore the
possibility of monopole or quadrupole modes, and the dimensionless
gaussian widths are taken as one. The same trial function,
Eq.~(\ref{trial}), also serves to characterize the stability of
the vortex in a small condensate~\cite{LF99}.

In evaluating the integration in $L(\Psi_v)$, we retain all terms
up to second order in the time-dependent parameters, yielding
\begin{eqnarray}
L_{\rm eff} &=& -2-\frac{\lambda}{2}-\frac{\gamma}{2\sqrt{2 \pi}}
+ \Omega \, \nonumber \\ &-&
\dot{\alpha}_x(2x_1-x_0)-\dot{\alpha}_y(2 y_1-y_0) \nonumber \\
&-&\dot{x}_0(y_1-y_0)+\dot{y}_0(x_1-x_0) \nonumber\\ &-&
\frac{1}{2}(x_1^2 + y_1^2)-x_0(x_1-x_0) -y_0(y_1-y_0) \nonumber \\
&-& \frac{1}{2}\left(\alpha_x^2 +\alpha_y^2 \right) +
\alpha_x(y_1-y_0)-\alpha_y(x_1-x_0)\nonumber
\\ &+&\Omega [x_0(x_1-x_0) + y_0(y_1-y_0) \nonumber \\
&-&\alpha_x (2 y_1-y_0) + \alpha_y(2 x_1-x_0)] \nonumber \\ &-&
\frac{\gamma}{\sqrt{2 \pi}} \left[(x_1-x_0)^2+ (y_1-y_0)^2 \right]
\end{eqnarray}
This Lagrangian takes a simpler form when written in terms of new
variables $\bbox{\delta} = \bbox{r}_1-\bbox{r}_0$ and
$\bbox{\epsilon} = 2 \bbox{r}_1-\bbox{r}_0$:
\begin{eqnarray}
L_{\rm eff} &=&-2-\frac{\lambda}{2}-\frac{\gamma}{2\sqrt{2 \pi}}+
\Omega \nonumber \\ &-&
\dot{\alpha}_x\epsilon_x-\dot{\alpha}_y\epsilon_y
-\left(\dot{\epsilon}_x-2\dot{\delta}_x\right)\delta_y
+\left(\dot{\epsilon}_y-2\dot{\delta}_y\right)\delta_x \nonumber
\\
&-&{1\over 2}\alpha^2 +\alpha_x\,(\delta_y-\Omega\epsilon_y)
+\alpha_y\,(-\delta_x+\Omega\epsilon_x) \nonumber \\
&-&{1\over2}\epsilon^2 +\Omega \,
\bbox{\delta}\cdot\bbox{\epsilon}
+\bigg(-2\Omega+{3\over2}-{\gamma \over \sqrt{2 \pi}}
\bigg)\delta^2 \, .
\end{eqnarray}

\section{Normal modes}

The effective Lagrangian leads to six coupled differential
equations for the displacements and the induced velocity.
Elimination of the latter through
\begin{equation} \label{alpha}
\bbox{\alpha} = \dot{\bbox{\epsilon}} + \hat{\bbox{z}} \times
(\Omega \, \bbox{\epsilon} - \bbox{\delta})
\end{equation}
yields two uncoupled pairs of homogeneous equations for the
displacements $\bbox{\delta}$ and $\bbox{\epsilon}$:
\begin{eqnarray}
\left(1 -{\gamma \over {2\sqrt{2\pi}}}-\Omega
\right)\delta_y(t)+\dot{\delta}_x(t)&=&0 \, ,
\\-\left(1 - {\gamma \over {2\sqrt{2 \pi}}}-\Omega \right)\delta_x(t)
+\dot{\delta}_y(t)&=&0 \, ,\\
 (\Omega^2-1)\epsilon_x(t)+2\Omega \dot{\epsilon}_y(t)
 - \ddot{\epsilon}_x(t) &=&0 \, \label{ex} ,\\ (\Omega^2 -1
)\epsilon_y(t)-2\Omega\dot{\epsilon}_x(t)-\ddot{\epsilon}_y(t)
&=&0 \label{ey} \, .
\end{eqnarray}
The $\bbox{\delta}$ mode with $\bbox{\epsilon}=0$ will be seen to
correspond to the anomalous mode; it has $\bbox{r}_0 =2
\bbox{r}_1$ (so that the displacement of vortex is twice that of
the condensate). In contrast, the $\bbox{\epsilon}$ mode with
$\bbox{\delta} =0$ represents a rigid dipole oscillation with
$\bbox{r}_0 =\bbox{r}_1$ (so that the vortex and the condensate
move together).

The physically relevant solutions of the equation of motion must
represent quantum states with positive normalization. A convenient
condition to check this normalization is~\cite{SF98}
\begin{equation} \label{norm}
\int dV  \, i \left( {n'}^* \Phi'-{\Phi'}^* n'\right) =
\frac{\hbar}{M} \, ,
\end{equation}
where $n'$ and $\Phi'$ are the complex fluctuations of the
particle density and the velocity potential respectively. The
density fluctuations are obtained by expanding the density $n=
|\Psi_v|^2 = n_0 + \delta n$, where $n_0 = (r^2/\pi^{3/2})
\exp(-r^2)\exp(-z^2)$ is the unperturbed density for the
condensate with a vortex. We find
\begin{equation}
\delta n \approx 2 n_0 \, \bbox{r} \cdot \left(\bbox{r}_1
-\frac{\bbox{r}_0}{r^2} \right) \, ,
\end{equation}
and use of the normal-mode amplitudes yields the expression in
plane polar coordinates
\begin{eqnarray}
\delta n = 2 n_0& \{ & \cos{\varphi}[r(\epsilon_x -\delta_x)
-r^{-1}(\epsilon_x - 2 \delta_x)]\nonumber \\ &+&
\sin{\varphi}[r(\epsilon_y -\delta_y) -r^{-1}( \epsilon_y - 2
\delta_y )]\} \, .
\end{eqnarray}
Given the explicit form of the displacements $\bbox{\delta}$ and
$\bbox{\epsilon}$ (see below), this real expression can be
converted to complex density fluctuations through
\begin{equation}
\delta n =n' e^{-i\omega t} + {n'}^* e^{i \omega t} \, .
\end{equation}

The complex fluctuations in the velocity potential follow by
expanding the phase of the condensate wave function $S =
\arctan[(y -y_0)/(x-x_0)] + \bbox{\alpha} \cdot \bbox{r} \approx
\varphi + \delta S$, since
\begin{equation}
\delta \Phi = (\hbar /M) \delta S = \Phi' e^{-i \omega t} +
{\Phi'}^* e^{i \omega t} \, .
\end{equation}
The vortex part $\delta S_v$ of this phase perturbation is $\delta
S_v = (-y_0 x + x_0 y)/r^2$, and condensate-velocity part $\delta
S_\alpha = \bbox{\alpha} \cdot \bbox{r} $ can be rewritten with
Eqs.~(\ref{alpha}). In terms of the normal mode variables
$\bbox{\delta}$ and $\bbox{\epsilon}$, we find
\begin{eqnarray}
\delta S &=& \cos{\varphi} \left[r
\dot{\epsilon}_x-\epsilon_y\left({1 \over r} + r \Omega \right)+
\delta_y \left( {2\over r} +r\right) \right] \nonumber
\\ &+& \sin{\varphi} \left[r \dot{\epsilon}_y
+\epsilon_x\left({1 \over r} + r \Omega \right)- \delta_x \left(
{2\over r} +r\right) \right] \, ,
\end{eqnarray}
and this expression can readily be converted into the complex
velocity potential $\Phi'$.

The equations of motion for the anomalous mode $\bbox{\delta}$
involve the characteristic frequency
\begin{equation}\label{omegaa}
\omega_a =-1 + {\gamma \over {2 \sqrt{2 \pi}}}+ \Omega  \, ,
\end{equation}
and the solution has the form
\begin{equation} \label{dt}
\bbox{\delta} (t) = \frac{1}{\sqrt{2}}\delta_0\left(\cos \omega_a
t \, , -\sin \omega_a t \right) \, ,
\end{equation}
where $\delta_0$ is an infinitesimal amplitude and the factor
$1/\sqrt{2}$ fixes the normalization of the perturbation
amplitudes (see below). Since $\bbox{\epsilon} =0$  for this mode,
the real small perturbation in the density becomes
\begin{equation}
\delta n_a= -2 n_0 \left(r -{ 2 \over r} \right)(\cos{\varphi} \,
\delta_x + \sin{\varphi} \, \delta_y ) \, ,
\end{equation}
and the complex density fluctuation (with time dependence $e^{-i
\omega_a t}$) is
\begin{eqnarray}
n'_a &=& -\frac{\delta_0}{\sqrt{2}} n_0 \left(r -
\frac{2}{r}\right)e^{-i \varphi} \, .
\end{eqnarray}
As expected from the physical picture of a single-particle
transition from the vortex condensate with unit angular momentum
per particle to the unoccupied axisymmetric gaussian ground state,
the quantum number of the $z$-component of the angular momentum in
this mode is $m=-1$. The real phase fluctuation for the anomalous
mode is
\begin{equation}
\delta S_a = \left( {2 \over r} +r \right)\left( \cos{\varphi} \,
\delta_y - \sin{\varphi} \, \delta_x\right)\, ,
\end{equation}
with the corresponding complex velocity potential
\begin{eqnarray}
{\Phi'_a}& =&  {\delta_0 \, \hbar \over 2\sqrt{2}iM}\left( {2
\over r} + r \right) e^{-i \varphi} \, .
\end{eqnarray}
For this anomalous mode, it is easy to verify that the
normalization integral~(\ref{norm}) has the proper positive value.
Since $\bbox{\epsilon} =0$, Eq.~(\ref{dt}) shows that the
amplitudes of the vortex $\bbox{r}_0$ and the condensate
$\bbox{r}_1$ become
\begin{eqnarray}
\bbox{r}_0 &=& \sqrt{2}\, \delta_0  \left(-\cos \omega_a t\, ,
\sin \omega_a t \right) \, , \\ \bbox{r}_1 &=& {1 \over 2}
\sqrt{2} \, \delta_0 \left(-\cos \omega_a t\, , \sin \omega_a t
\right) \, ;
\end{eqnarray}
they both execute small-amplitude in-phase motion with
$\bbox{r}_0$ twice as large as $\bbox{r}_1$. For $\Omega =0$, the
anomalous frequency (\ref{omegaa}) is negative, and the motion is
counter-clockwise in a positive direction (namely right-circular
motion with positive helicity). With increasing rotation frequency
$\Omega$,  the negative anomalous frequency increases towards zero
and vanishes at the metastable rotation frequency $\Omega_m =
1-\gamma/(2 \sqrt{2 \pi})$~\cite{LF99}. For $\Omega > \Omega_m$,
the apparent motion becomes clockwise (namely left-circular motion
in a negative direction). The expressions found here for the
density and velocity potential fluctuations $n'_a$ and $\Phi'_a$
coincide with those found from the Bogoliubov
approach~\cite{Fet98}. If we had omitted either the induced
velocity or the displacement of the condensate, we would have
found different frequencies, just as in the discussion of vortex
stability~\cite{LF99}. The present Lagrangian treatment has the
advantage of providing a physical picture of the motion,
clarifying the nature of the anomalous mode. In the weak-coupling
limit, the vortex and condensate move in phase, in contrast to the
behavior found in the strong-coupling (Thomas-Fermi)
limit~\cite{Svid98}.

In a similar manner, we can treat the dipole modes, which have
$\bbox{\delta} =0$, so that $\bbox{r}_0=\bbox{r}_1$ and the vortex
moves rigidly with the condensate. Equations (\ref{ex}) and
(\ref{ey}) show that the $x$ and $y$ motions are uncoupled if
$\Omega =0$. In general, however, there are two coupled normal
modes with frequencies $\omega_+ = 1-\Omega$ and $\omega_- = 1 +
\Omega$. The notation $\omega_\pm$ reflects the general
construction of the Bogoliubov amplitudes $u_\pm = a^{\dagger}_\pm
\Psi$ and $v_\pm = a_\mp \Psi^*$ for the dipole modes in the
laboratory frame, where $\Psi$ is any solution of the
Gross-Pitaevskii equation in an axisymmetric trap; the resulting
density perturbation is $n'_\pm \propto e^{\pm i \varphi} e^{-i
\omega_\pm t}$, see Eqs.~(\ref{npm}) and (\ref{phipm}) below. It
is easy to see that the general solution is a linear combination
of the two distinct normal modes
\begin{eqnarray}
\bbox{\epsilon}_{+} (t) &=& \epsilon_{+}(0) \left(\cos{\omega_+ t}
\, ,\sin{\omega_+ t}\right) \, ,
\\ \bbox{\epsilon}_{-} (t) &=& \epsilon_{-}(0)
\left(\cos{\omega_- t} \, ,- \sin{\omega_- t}\right)  \, ,
\end{eqnarray}
where $\epsilon_{+}(0),\epsilon_{-}(0)$ are infinitesimal small
amplitudes.

We can now check the normalization for the dipole modes. The
first-order contribution to the density for both modes is given by
\begin{equation}
\delta n_{\pm} = 2 n_0 \left(r -{1 \over r} \right)(\cos{\varphi}
\, \epsilon_{x \pm} +\sin{\varphi} \, \epsilon_{y \pm}) \, ,
\end{equation}
and hence the complex density fluctuations are
\begin{eqnarray} \label{npm}
n'_{\pm} &=& \epsilon_\pm (0) \, n_0 \left(r -{1\over r}
\right)e^{\pm i \varphi} \, .
\end{eqnarray}
For the phase fluctuations we find
\begin{equation}
\delta S_{\pm} = \left[ \frac{1}{r}+ r (\Omega \pm \omega_{\pm})
\right](\sin {\varphi}\, \epsilon_x - \cos{\varphi}\, \epsilon_y)
\, ,
\end{equation}
and the resulting complex fluctuations for the velocity potential
are
\begin{eqnarray} \label{phipm}
\Phi'_{\pm}&=& {\epsilon_\pm (0) \, \hbar \over 2iM}\left( r \pm
{1 \over r } \right) e^{\pm i \varphi} \, ,
\end{eqnarray}
independent of the external rotation frequency $\Omega$. With
these expressions, the normalization integral is one for both
normal modes with frequencies $\omega_{\pm}$.

The final expression for the motion of the vortex and the
condensate in the dipole modes [$\bbox{r} (t) \equiv \bbox{r}_0
(t) =\bbox{r}_1(t)$] is
\begin{eqnarray}
\bbox{r}_\pm(t) &=& r_0\left(\cos{\omega_\pm t} \, , \,
 \pm \sin{\omega_\pm t}\right)\, ,
\end{eqnarray}
with an infinitesimal amplitude $r_0$. We again find circular
trajectories, and here they are the same for the vortex and the
condensate motion, confirming that the modes are indeed the rigid
dipole oscillations of the vortex and the condensate. For
stability reasons, only the region $ \mid \Omega \mid <1$ is
relevant; thus the motion is always counter-clockwise (positive)
for $\omega_+$ and clockwise (negative) for $\omega_-$. Comparison
with the Bogoliubov results~\cite{Fet98} shows exact agreement for
$\Omega =0$. The rotation splits the dipole frequencies by twice
the external rotation frequency. This result can be visualized by
expressing the dipole modes (for $\Omega =0$) in terms of circular
quanta, namely as circular motions in the positive and negative
sense. For zero external rotation, the angular velocities are
degenerate. In a frame rotating with $\Omega$ in the positive
sense, the positive mode rotates at the difference  $1 -\Omega$
and hence slows down. The negative mode, however, speeds up by the
same amount, leading to an effective splitting in the rotating
frame of $ 2 \Omega$.

\section{conclusion}

We have shown that a time-dependent Lagrangian approach provides
physical insight into the nature of the low-energy modes of a
vortex in a small trapped Bose-Einstein condensate. The anomalous
mode is an in-phase circular motion of the vortex and the
condensate with different amplitudes. In a rotating frame, the
sense of this mode reverses from positive to negative when the
external rotation frequency exceeds the metastable
frequency~\cite{LF99}. The dipole modes are rigid rotations of the
vortex and the condensate together. In terms of the density and
phase fluctuations, all these results agree exactly with those
from the Bogoliubov approach~\cite{Fet98}.

\acknowledgements This work was supported in part by NSF Grant
No.~94-21888 and by the DAAD (German Academic Exchange Service)
``Doktorandenstipendium im Rahmen des gemeinsamen
Hochschulsonderprogramms III von Bund und L\"andern" (M.~L.). A.~L.~F.
thanks the Aspen Center for Physics where part of this work was 
performed.


\begin{references}



\bibitem{And} M.~H.~Anderson, J.~R.~Ensher, M.~R.~Matthews,
C.~E.~Wieman, and E.~A.~Cornell, Science~{\bf 269}, 198 (1995).



\bibitem{Dav}  K.~B.~Davis, M.-O.~Mewes, M.~R.~Andrews, N.~J.~van
Druten, D.~S.~Durfee, D.~M.~Kurn, and W.~Ketterle,
Phys.~Rev.~Lett.~{\bf 75}, 3969 (1995).



\bibitem{Brad}C.~C.~Bradley, C.~A.~Sackett, and R.~G.~Hulet,
Phys.~Rev.~Lett.~{\bf 78}, 985 (1997).

\bibitem{B96} G.~Baym and C.~J.~Pethick, Phys.~Rev.~Lett. {\bf
76}, 6 (1996).

\bibitem{Dodd} R.~J.~Dodd, K.~Burnett, M.~Edwards, and C.~W.~Clark,
Phys.~Rev.~A {\bf 56}, 587 (1997).

\bibitem{Fet98} A.~L.~Fetter, J.~Low~Temp.~Phys. {\bf 113}, 198
(1998).

\bibitem{SF98}  A.~A.~Svidzinsky and A.~L.~Fetter, Phys.~Rev.~A {\bf
58}, 3168 (1998).

\bibitem{Svid98} A.~A.~Svidzinsky and A.~L.~Fetter,
cond-mat/9811348.

\bibitem{Zam98} F.~Zambelli and S. Stringari, Phys.~Rev.~Lett.
{\bf 81}, 1754 (1998).

\bibitem{IsoJ99} T.~Isoshima and K.~Machida, J.~Phys.~Soc.~Jpn.
{\bf 68}, 487 (1999).

\bibitem{Ga99} J.~J.~Garc\'{\i}a-Ripoll and
V.~M.~P\'{e}rez-Garc\'{\i}a, cond-mat/9903353.

\bibitem{Bog47} N.~Bogoliubov, J.~Phys.~(Moscow) {\bf 11}, 23
(1947).

\bibitem{Pe97} V.~M.~P\'{e}rez-Garc\'{\i}a, H.~Michinel,
J.~I.~Cirac, M.~Lewenstein and P.~Zoller, Phys.~Rev.~A {\bf 56},
1424 (1997).


\bibitem{LF99} M.~Linn and A.~L.~Fetter, cond-mat/9906139.

\end{references}
\end{document}